# QTPIE: Charge Transfer with Polarization Current Equalization. A fluctuating charge model with correct asymptotics

Jiahao Chen and Todd J. Martínez*
Department of Chemistry and the Beckman Institute
University of Illinois
Urbana, Illinois 61801

**Abstract.** Polarization and charge transfer are important effects which are difficult to describe using conventional force fields. Charge equilibration models can include both of these effects in large-scale molecular simulations. However, these models behave incorrectly when bonds are broken, making it difficult to use them in the context of reactive force fields. We develop a new method for describing charge flow in molecules – QTPIE. The QTPIE method is based on charge transfer variables (as opposed to atomic charges) and correctly treats asymptotic behavior near dissociation. It is also able to provide a realistic description of in-plane polarizabilities.

---

*Address: 600 S. Mathews Ave., Urbana IL 61801 E-mail: tjm@spawn.scs.uiuc.edu

**Introduction**

Polarization and charge transfer effects are known to be important components of molecular interactions. However, these effects are difficult to model correctly in the context of empirical force fields which are applicable to large scale simulations. A promising approach is based on the Drude oscillator, where each atom has a charge attached to it through a harmonic spring [1-3]. However, these "charge-on-spring" or "shell" models cannot describe charge transfer, which may be viewed as an extreme manifestation of polarization. Other approaches based on atom-centered multipole expansions suffer the same restriction to polarization in practice, because the multipole expansion is normally truncated at the dipole term [4,5]. Ideally, one would use a model which makes no artificial distinction between charge transfer and polarization. The most promising methods which have been proposed in this context are the "fluctuating charge" (also called charge equilibration or chemical potential equalization) models [6,7], based on the chemical concepts of electronegativity and hardness. The basic idea has a long history, tracing back to the introduction of the concept of electronegativity by Pauling [8], through a variety of highly parameterized models [9-11] and culminating with the minimally-parameterized QEq [6] and *fluc-q* [7,12] methods. However, there are difficulties [13,14] in reconciling electronegativity concepts with the known discontinuities of the derivative of the electronic energy with respect to the number of electrons [15]. We have previously analyzed charge equilibration methods from a wavefunction viewpoint in order to clarify some of the important issues [16,17]. Two important ideas arising from that work were the identification of charge transfer as the fundamental variables (in place of atomic charges) and the need for a "pairwise electronegativity," dependent on the distance between any pair of atomic centers. In this work, we take those ideas a step further to develop a charge equilibration method based on charge transfer variables and distance-dependent electronegativity.

We begin with a brief overview of charge equilibration methods, using the QEq method as an example. In QEq, the electrostatic energy of a molecular system is given by pairwise Coulomb interactions plus an internal energy term expanded to second order in the partial charges:

$$E(\mathbf{q}) = \sum_{i=1}^{n}\left(\chi_i^0 q_i + \frac{1}{2}\eta_i q_i^2\right) + \sum_{i<j} q_i q_j J_{ij} \qquad (1)$$

where $i$ indexes the $n$ atomic sites. The screened Coulomb interaction $J_{ij}$ may be represented as an integral over single $ns$-type Slater orbitals (STOs):

$$J_{ij}(\mathbf{R}_{ij}) = \left\langle \phi_i \phi_j \left| \frac{1}{|\mathbf{r}_i - \mathbf{r}_j|} \right| \phi_i \phi_j \right\rangle; \quad \phi_i(\mathbf{r}_i; \mathbf{R}_i) = N_i |\mathbf{r}_i - \mathbf{R}_i|^{n-1} e^{-\zeta_i |\mathbf{r}_i - \mathbf{R}_i|} \qquad (2)$$

where $\mathbf{r}_i$ and $\mathbf{R}_i$ refer to the positions of electrons and nuclei, respectively, and $\mathbf{R}_{ij}$ is the distance between the $i$th and $j$th atoms. The coefficients in the first two terms of the expansion are identified as Mulliken electronegativities [18], $\chi_i^0$, and Parr-Pearson hardnesses [19,20], $\eta_i$:

$$\begin{aligned}\chi_i^0 &= \frac{1}{2}(\mathrm{IP}_i + \mathrm{EA}_i) = -\mu_i \\ \eta_i &= \frac{1}{2}(\mathrm{IP}_i - \mathrm{EA}_i)\end{aligned} \qquad (3)$$

where $\mathrm{IP}_i$ and $\mathrm{EA}_i$ are the ionization potential and electron affinity of the $i$th atom, respectively, and we have noted that the electronegativity $\chi_i^0$ is trivially related to an atomic chemical potential $\mu_i$. The QEq partial charges $q_i$ are obtained by minimizing the energy expression of Eq. (1) under the constraint of fixed total number of electrons. This can be expressed as a linear system of equations, solved by inverting an $n \times n$ matrix.

The QEq model has been shown to work well for chemically reasonable structures near equilibrium [6]. However, it suffers from some fundamental problems that limit its utility in reaction dynamics. One of the most important is the unreasonable charge distributions predicted for geometries far from equilibrium, which has serious implications for its use in reactive molecular dynamics simulations involving bond dissociation [14,16,17]. For example, the QEq solution for a neutral diatomic molecule is

$$q_2 = -q_1 = \frac{\chi_2^0 - \chi_1^0}{J_{11} - 2J_{12} + J_{22}} \Rightarrow \lim_{R_{12} \to \infty} q_2 = \frac{\chi_1^0 - \chi_2^0}{J_{11} + J_{22}} \neq 0 \qquad (4)$$

The partial charges do not vanish in the asymptotic limit $R_{12} = |\mathbf{R}_1 - \mathbf{R}_2| \to \infty$ and are in general some non-integral value, which is unphysical. In general, QEq and similar charge equilibration models overestimate the propensity for charge flow in polyatomic molecules, giving rise to inflated values of molecular electrostatic properties such as dipole moments and polarizabilities, especially for geometries far from equilibrium [21]. We therefore desire a fluctuating-charge model that can predict partial charges in such geometries with at least qualitative accuracy.

**Theory**

We have previously analyzed the behavior of charge equilibration methods in detail and proposed an improved model which addresses their shortcomings [16,17]. The equations defining the proposed model were written explicitly and tested numerically for a diatomic molecule. Here, we generalize the relevant equations to polyatomic molecules and carry out some tests of the method's numerical accuracy. Our new method shifts the focus away from the atomic partial charges **q** onto charge transfer variables **p** that describe a polarization current, i.e. a tendency for electronic density to migrate from one atom onto another. The method is thus named QTPIE, for charge transfer with polarization current equilibration. The charge transfer variables are related to the atomic charges by continuity:

$$q_i = \sum_j p_{ji} \qquad (5)$$

where $p_{ji}$ describes the amount of charge transferred from the $i$th atom to the $j$th atom. By symmetry, the charge transfer variables must form an antisymmetric matrix, i.e. $p_{ij} = -p_{ji}$. A similar variable transformation has been previously introduced to simplify the numerical solution of charge equilibration equations [22]. While the variable transformation alone does nothing to change the results of the method, it does suggest an improved model. In terms of charge transfer variables, the QEq energy expression has the form:

$$E(\mathbf{p}) = \sum_{ij} \chi_i^0 p_{ji} + \frac{1}{2} \sum_{ijkl} p_{ki} p_{lj} J_{ij} \qquad (6)$$

where we have defined $J_{ii} = \eta_i$. Minimization of this expression with respect to the charge transfer variables will lead in general to unphysical finite charge transfer between infinitely separated atoms. Thus, we introduce a distance-dependent function which penalizes long-range charge transfer to obtain a generalized energy expression, which is the central equation of QTPIE:

$$\begin{aligned} E(\mathbf{p}) &= \sum_{ij} \chi_i^0 f_{ji} p_{ji} + \frac{1}{2} \sum_{ijkl} p_{ki} p_{lj} J_{ij} \\ &= \sum_{i<j} p_{ji} \left[ \left( \chi_j^0 - \chi_i^0 \right) f_{ji} + \frac{1}{2} \sum_{k<l} p_{lk} \left( J_{ik} - J_{jk} - J_{il} + J_{jl} \right) \right] \end{aligned} \qquad (7)$$

As shown previously [16], the function $f_{ji}$ should decay with distance (on a length scale related to the orbitals involved on atoms $i$ and $j$). By detailed balance, it should also be invariant to index exchange, i.e. $f_{ij}=f_{ji}$. The simplest choice of $f_{ij}$ is therefore an overlap integral between orbitals on the $i$th and $j$th atoms. In the present work, we take this function to be a scaled overlap integral of the n$s$-type orbitals which are used to represent the screened Coulomb interaction, i.e.

$$f_{ji} = k_{ji} S_{ji} = k_{ji} \langle \phi_j | \phi_i \rangle \qquad (8)$$

The scaling factor $k_{ji}$ could be optimized, but for most of the present work we simply choose $k_{ji}$ to be unity for all atom pairs. The sum in Eq. (7) is *not* limited to atom pairs which are involved in covalent bonds, and thus there is no need for *a priori* determination of what atoms are bonded. We use the QEq parameters for electronegativities, hardnesses, and orbital radii without modification. Explicit reparameterization can thus be expected to improve all of the results reported here. We note that, in principle, one could optimize the parameter $k_{ij}$ for different bond types, e.g. aromatic C-C, but the results presented here suggest that this will not be necessary. In the second line of Eq. (7), we exploit the antisymmetric nature of the charge transfer variables and the symmetric nature of $f_{ij}$ to write the equation in skew-symmetric form.

Requiring the energy of Eq. (7) to be minimized with respect to all charge transfer variables leads to the system of linear simultaneous equations

$$\forall i,j: 0 = \frac{\partial E}{\partial p_{ji}} = \left(\chi_j^0 - \chi_i^0\right)k_{ji}S_{ji} + \sum_{k<l} p_{lk}\left(J_{ik} - J_{jk} - J_{il} + J_{jl}\right) \tag{9}$$

The QTPIE solution for a diatomic molecule is thus:

$$q_2 = p_{21} = \frac{\chi_2^0 - \chi_1^0}{J_{11} - 2J_{12} + J_{22}} k_{12}S_{12} \Rightarrow \lim_{R_{12} \to \infty} q_2 = 0 \tag{10}$$

In contrast to the QEq solution of Eq. (4), QTPIE correctly predicts vanishing charge transfer in the dissociation limit and should therefore provide a more accurate description of fluctuating charges at non-equilibrium geometries such as those along reactive trajectories.

**Results and Discussion**

The QEq and QTPIE methods were implemented in Scilab [23] and solved in a linear algebraic representation in the space of unique atomic pairs. We did not implement the charge-dependent atomic radius for hydrogen atom described in the original QEq method, but instead use the equations as embodied in Eqs. (1)-(3). Thus, the results presented here are denoted QEq(-H), indicating that the hydrogen correction is not employed. For the QTPIE method, the linear system of Eq. (9) is highly singular, implying that the number of independent variables is far less than the number of charge transfer variables ($\approx n^2$, where $n$ is the number of atoms). In fact, the number of independent variables is exactly the same in the two methods, i.e. ($n$-$1$). This equivalence arises because charge transfer around closed loops does not influence the energy expression of Eq. (7). Singular value decomposition [24] was used to construct the pseudoinverse in the solution of the linear equations for QTPIE. As discussed above, we made no attempt to reparameterize the QTPIE method. Instead, we use the parameters previously optimized for QEq [6], except where explicitly stated otherwise. As in QEq, all parameters are dependent only on the identity of a given atom. In other words, there is no attempt to establish different parameters for $sp^2$ and $sp^3$ carbon atoms, for example. We do not introduce any *a priori* information about the covalent bonds in the molecule – all information about molecular connectivity is embedded in the screened Coulomb interaction and the attenuation factor $f_{ij}$.

We performed calculations on three representative small molecules: sodium chloride, water and phenol. For each molecule, we compare the predictions of QEq(-H) and QTPIE with the results of *ab initio* calculations. Since atomic charges are not well-defined quantum-mechanical observables, we chose several distinct definitions for comparison, namely Mulliken population [25] and distributed multipole [26] analysis (DMA). The DMA calculation was restricted to monopoles on the atomic centers. The electronic structure calculations for these charge analyses were in general performed using multi-reference *ab initio* methods with small basis sets. We choose small basis sets with limited spatial extent so as to facilitate comparisons between the *ab initio* and QTPIE/QEq methods, since both QTPIE and QEq(-H) use a minimal basis set representation of the atomic charge density in the screened Coulomb interaction.

For illustrative purposes, we present results of the QEq(-H) and QTPIE models applied to isolated sodium chloride molecule at different internuclear distances. *Ab initio* results are obtained from a complete active space (CAS) calculation [27] using eight electrons in five orbitals, i.e. CAS(8/5), with a 3-21G basis set [28]. This full valence active space wavefunction is able to describe both ionic and covalent character. Because of the weakly avoided crossing between the covalent and ionic diabatic states, the transition from ionic to covalent character on the ground electronic state is quite rapid, as seen in both the Mulliken and DMA charges shown in Figure 1. The Mulliken and DMA definitions of the atomic charges give similar values throughout, indicating the robustness of the *ab initio* partial charges we are using for comparison with the QTPIE and QEq(-H) results. Figure 1 shows that, as expected from Eq. (4), the QEq(-H) method predicts finite charge transfer at infinite separation: asymptotically $q_{Na} = -q_{Cl} = 0.394$. However, QTPIE correctly predicts that there is no charge transfer in the dissociated molecule. The QTPIE charges are not in quantitative agreement with the *ab initio* charges. This is expected, given the weakly avoided crossing (at large internuclear distance) between the covalent and ionic states in this molecule. Only a fully quantum mechanical method could be expected to predict the rapid transition between covalent and ionic states correctly.

Since the dissociation catastrophe and the ensuing divergence of dipole moments is one of the most serious limitations of QEq and other charge equilibration models, we also investigate the behavior of partial charges in an asymmetrically dissociated water molecule. In this hypothetical reaction, the H-O-H internal bond angle was set to $\theta = 104.5°$ and one of the O-H bonds was kept fixed at 0.97Å while the other O-H bond length was varied. The *ab initio* data were computed at the CAS(10,7)/STO-3G level of theory.

In Figure 2, we show the atomic charges on the dissociating hydrogen and oxygen atom computed from *ab initio,* QEq(-H), and QTPIE methods. The atomic charge on the remaining hydrogen atom can be deduced by considering overall charge neutrality. Similar to the NaCl example, the QEq(-H) values do not exhibit the correct asymptotics. In contrast, the QTPIE charge on the dissociating hydrogen atom vanishes correctly in the limit of infinite separation. The QTPIE partial charge on the oxygen atom in the OH fragment is closer to the *ab initio* result than that predicted by QEq(-H). However, it is still too large, indicating an overestimation of the dipole moment of OH. Thus, we attempt the simplest reparameterization possible, namely varying $k_{OH}$ of Eq. (8), while demanding that $k_{OH}=k_{HH}$. We chose the value for $k_{OH}$ which led to agreement of the partial charge on oxygen atom at the equilibrium geometry of the water molecule ($k_{OH}=k_{HH}=0.4072$). With this modification, the QTPIE charges are in good agreement with the *ab initio* values across the whole range of O-H distances, as shown in Figure 3.

In order to explore the adequacy of a single set of QTPIE parameters for other molecular geometries, we computed similar dissociation curves with varying ∠HOH in the range 60°-150°. The results (using $k_{OH}=k_{HH}=0.4072$, as discussed above) are compared with *ab initio* charges from Mulliken analysis on CASPT2(10,7)/STO-3G data in Figure 4. The results from QTPIE remain in similarly good agreement with the *ab initio* calculations for all of these geometries, particularly in the dissociation limit.

It is important that a fluctuating charge model be able to accurately model the change in atomic charges with response to an external electric field. Thus, we have also computed the molecular polarizability tensor using QEq and QTPIE. These results are again compared with *ab initio* calculations.

The QEq model has two shortcomings when computing molecular polarizabilities. The first is a tendency to overestimate the in-plane components, which is in part due to the overestimation of charges for weakly interacting (i.e. widely separated) atoms. The second is its inability to calculate the out-of-plane component of the molecular polarizability tensor for planar molecules. This latter deficiency arises because the model considers only atomic charges and not atomic dipoles or charge centers apart from the locations of the atoms. This makes it impossible to have charge fluctuations along any direction other than in directions directly leading to another point charge. In terms of molecular graphs, charge flow is restricted only to edges and therefore cannot flow out of the plane of the molecule. Similar restrictions apply in the QTPIE method as described here, and thus one might expect that QTPIE will also fail to describe the out-of-plane polarizabilities for planar molecules. Extensions of charge equilibration methods which expand the basis which describes the charge fluctuations from a single *s* function per atomic site to include also *p*-type functions are a promising route to solve this problem [29]. We plan to investigate the application of similar ideas to QTPIE in future work.

The QTPIE energy expression in an external field $\vec{\varepsilon}$ is given by:

$$E(\mathbf{p};\vec{\varepsilon}) = \sum_{ij} \chi_i^0 f_{ji} p_{ji} + \frac{1}{2} \sum_{ijkl} p_{ki} p_{lj} J_{ij} + \sum_{ij} p_{ij} \vec{x}_i \cdot \vec{\varepsilon} \qquad (11)$$

We compute the QTPIE polarizability by finite differencing the dipole moment with respect to the external field, solving for the optimal charge transfer variables in Eq. (11) at each value of the external field. The scaling factor for the overlap, $k_{ij}$, was taken to be unity in all of these QTPIE calculations. Table 1 summarizes the results for sodium chloride, water and phenol. The *ab initio* polarizabilities were calculated as second derivatives of the second-order Møller-Plesset perturbation theory (MP2) energy using the method of finite fields. The *ab initio* calculations use an aug-cc-pVDZ basis set [30,31] which includes the diffuse functions necessary for accurate calculations of polarizabilities. Ground state equilibrium geometries were optimized using MP2/aug-cc-pVDZ and these geometries were used for QEq(-H), QTPIE, and *ab initio* calculations of the polarizability.

Molecular polarizabilities calculated using the three methods above were found to be stable with respect to small perturbations in the nuclear geometries, so discrepancies in the eigenvalues due to geometric effects can be ruled out. As expected, both QEq(-H) and QTPIE incorrectly predict a vanishing out-of-plane component of the polarizability for these planar molecules. The eigenvalues of the polarizability tensor in QTPIE are smaller than those from QEq(-H), generally by a factor of approximately two. The in-plane components of the polarizability predicted by QTPIE are in good agreement with the *ab initio* values.

**Conclusions**

We have defined a new fluctuating charge model (QTPIE), which defines atomic charges as sums over charge-transfer variables that describe polarization currents. This construction allows us to create a simple fluctuating-charge model that exhibits correct asymptotic behaviors for weakly-interacting atoms, i.e. near dissociation. We showed that the QTPIE model also describes molecular polarizabilities more accurately than previous fluctuating charge models. We did not make any significant attempt to optimize the parameters for QTPIE, but instead used parameters (electronegativities, hardnesses, and orbital radii for the shielded Coulomb interaction) optimized for the QEq method. One may expect improved results if the parameters are reoptimized for QTPIE, and this is currently under investigation. We hope that the successes and limitations of QTPIE will motivate the development of even more accurate fluctuating-charge models and improve our understanding of the chemically useful concepts of electronegativity and hardness.

**Acknowledgments.** This work was supported by NSF DMR-03 25939 ITR, via the Materials Computation Center at the University of Illinois at Urbana-Champaign, and by DOE DE-FG02-05ER46260. TJM is a MacArthur fellow.

**Table 1.** Eigenvalues (sorted by descending magnitude) of the dipole polarizability tensor (in units of Å$^3$) for three molecules.

|        | QEq(-H) | QTPIE | MP2/aug-cc-pVDZ |
|--------|---------|--------|-----------------|
|        | 13.9474 | 6.2171 | 4.5042 |
| NaCl   | 0.0000  | 0.0000 | 3.6932 |
|        | 0.0000  | 0.0000 | 3.6931 |
|        | 3.4653  | 1.8338 | 1.4502 |
| H$_2$O | 1.2317  | 0.6516 | 1.3678 |
|        | 0.0000  | 0.0000 | 1.2883 |
|        | 24.6244 | 13.0298 | 13.6758 |
| Phenol | 20.3270 | 10.7566 | 12.3621 |
|        | 0.0000  | 0.000  | 6.9981 |

**Figure Legends**

**Figure 1.** Partial charges (in atomic units) on dissociating NaCl as computed using QEq(-H) and QTPIE. *Ab initio* data from CAS(8,5)/3-21G calculations were also analyzed using Mulliken population and distributed monopole analysis (DMA). The QEq(-H) method predicts significant charge transfer at dissociation, while the QTPIE method predicts uncharged fragments in this limit, in agreement with the *ab initio* results. The experimentally-determined equilibrium bond length of NaCl is indicated on the graph ($R_{eq}$=2.361Å).

**Figure 2.** Partial charges (in atomic units) for a dissociating water molecule. Positive values are the charge on the dissociating hydrogen atom, and negative values correspond to charges on the oxygen atom. Distributed multipole analysis (DMA) charges are obtained from a CAS(10/7) wavefunction in a minimal basis set (STO-3G). The QTPIE method without any reparameterization reproduces the vanishing charge on the dissociating hydrogen atom at infinite separation predicted by the *ab initio* method.

**Figure 3.** As in Figure 2, but using $k_{OH}=k_{HH}=k$ of Eq. 8 which is optimized ($k$=0.4072) to give agreement of QTPIE and DMA charges at the equilibrium geometry of the water molecule. Even with this very limited reparameterization, the QTPIE method agrees well with *ab initio* charges throughout (except for very short bond distances, where the concept of partial charges becomes suspect due to linear dependence).

**Figure 4.** As in Figure 3, but for varying internal angles $\theta$.

**Figure 5.** Atomic partial charges for phenol in the equilibrium geometry computed with MP2/cc-pVDZ. Charges from QTPIE, QEq(-H) (in bold), and *ab initio* Mulliken analysis (in italic) are shown for each atom.

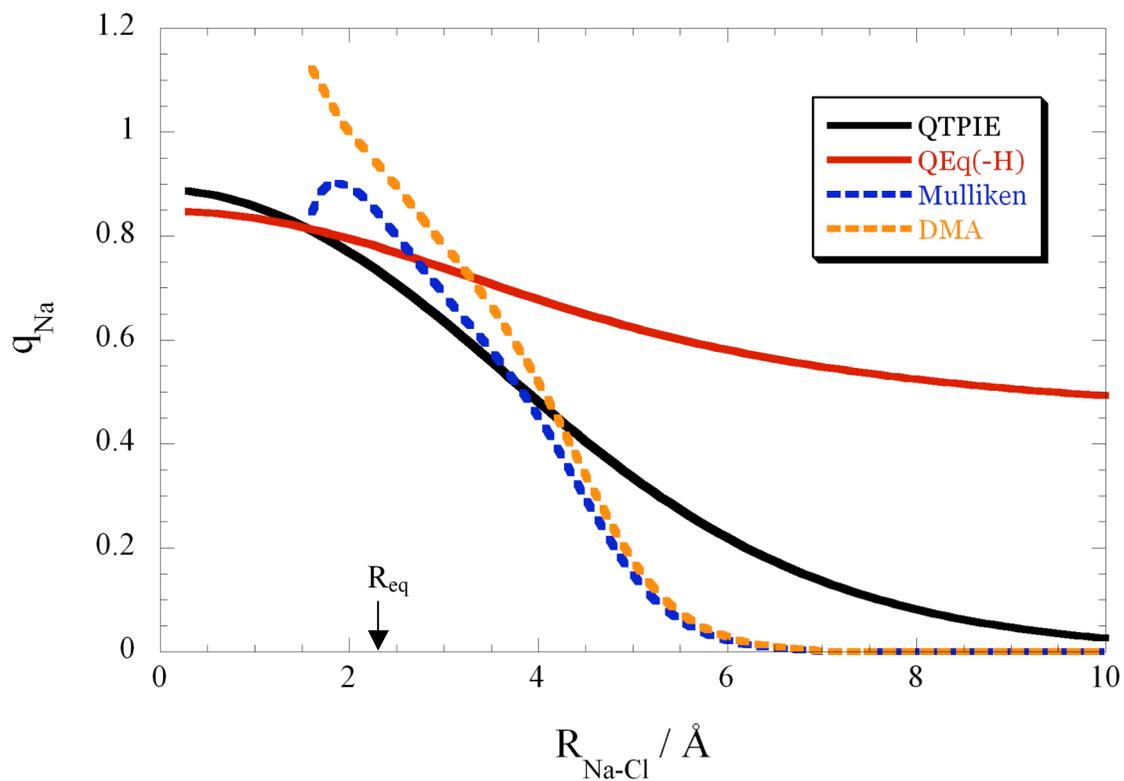

**Figure 1.** Partial charges (in atomic units) on dissociating NaCl as computed using QEq(-H) and QTPIE. *Ab initio* data from CAS(8,5)/3-21G calculations were also analyzed using Mulliken population and distributed monopole analysis (DMA). The QEq(-H) method predicts significant charge transfer at dissociation, while the QTPIE method predicts uncharged fragments in this limit, in agreement with the *ab initio* results. The experimentally-determined equilibrium bond length of NaCl is indicated on the graph ($R_{eq}$=2.361Å).

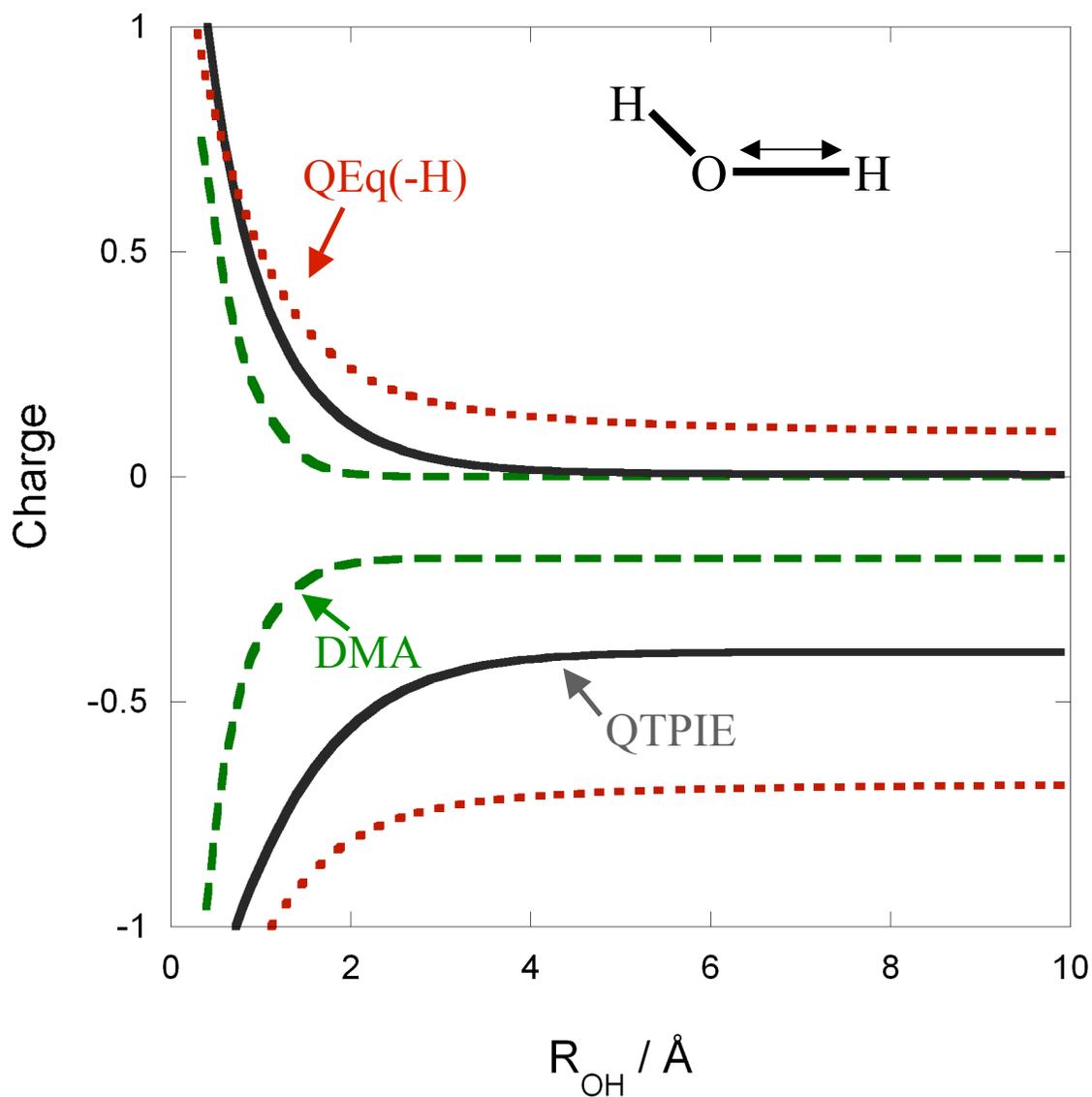

**Figure 2.** Partial charges (in atomic units) for a dissociating water molecule. Positive values are the charge on the dissociating hydrogen atom, and negative values correspond to charges on the oxygen atom. Distributed multipole analysis (DMA) charges are obtained from a CAS(10/7) wavefunction in a minimal basis set (STO-3G). The QTPIE method without any reparameterization reproduces the vanishing charge on the dissociating hydrogen atom at infinite separation predicted by the *ab initio* method.

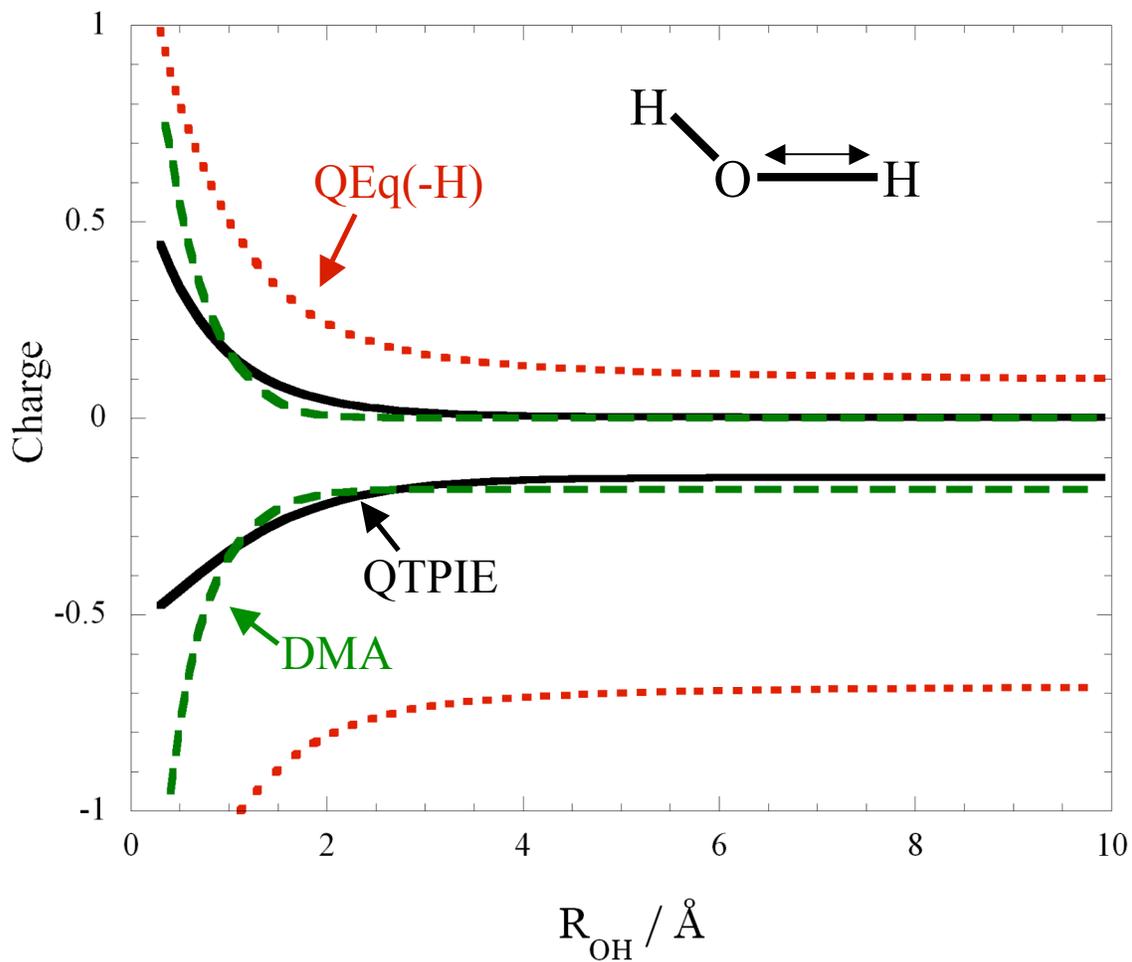

**Figure 3.** As in Figure 2, but using $k_{OH}=k_{HH}=k$ of Eq. 8 which is optimized ($k=0.4072$) to give agreement of QTPIE and DMA charges at the equilibrium geometry of the water molecule. Even with this very limited reparameterization, the QTPIE method agrees well with *ab initio* charges throughout (except for very short bond distances, where the concept of partial charges becomes suspect due to linear dependence).

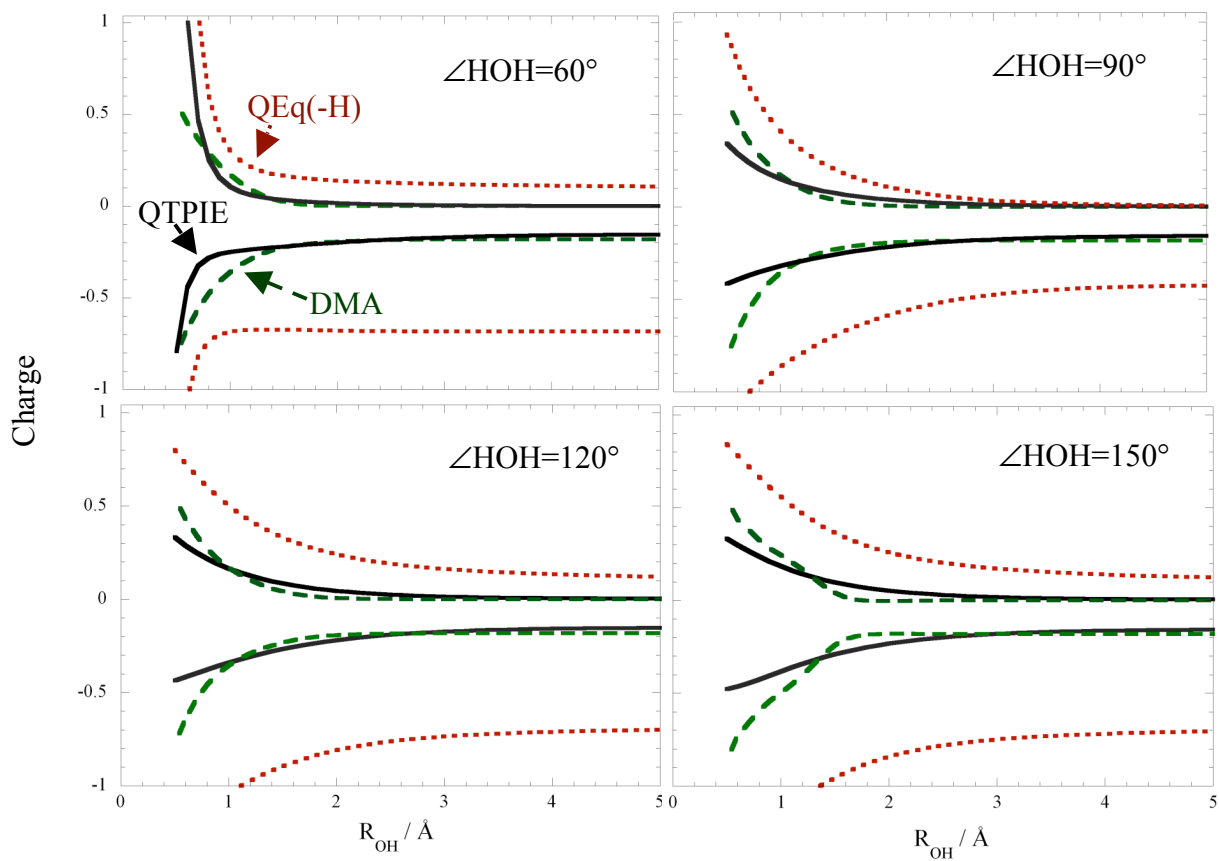

**Figure 4.** As in Figure 3, but for varying internal angles θ.

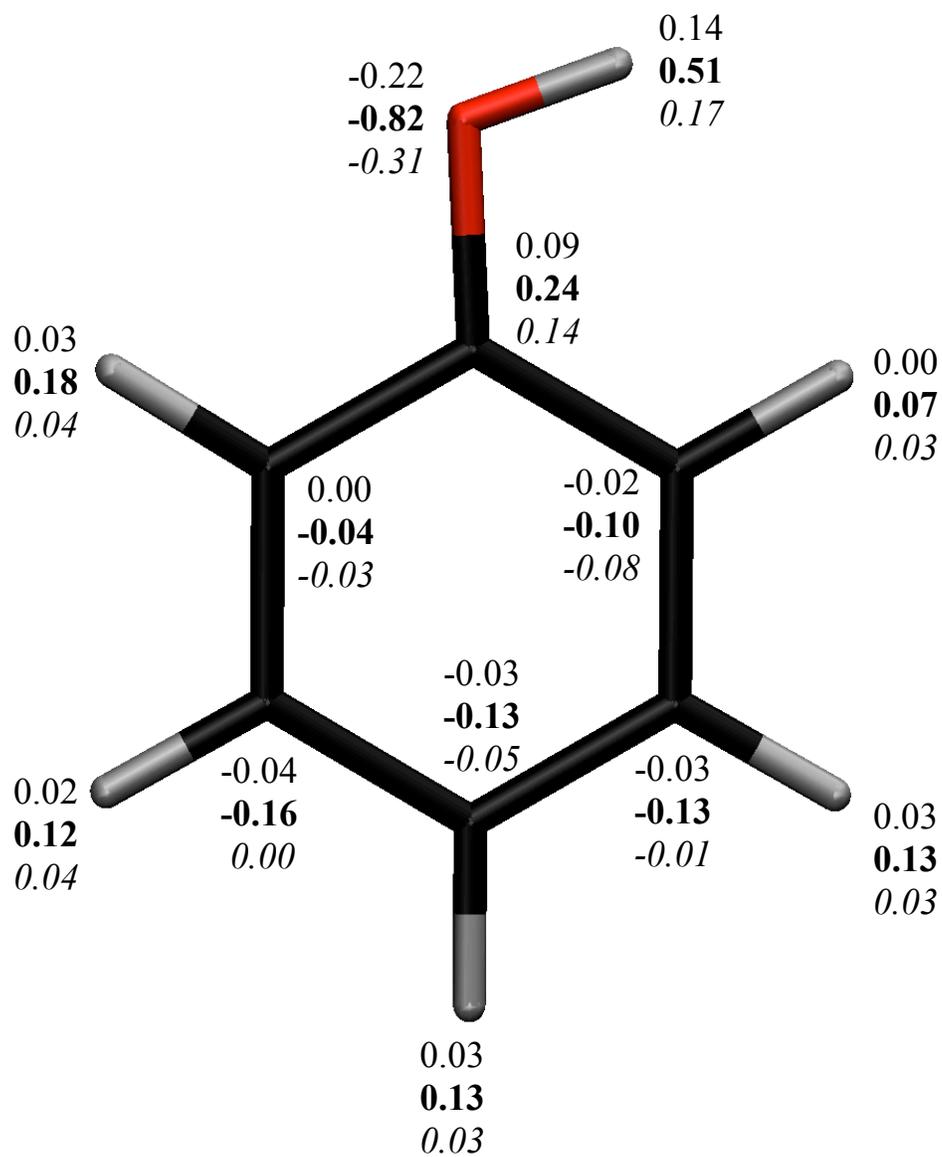

**Figure 5.** Atomic partial charges for phenol in the equilibrium geometry computed with MP2/cc-pVDZ. Charges from QTPIE, QEq(-H) (in bold), and *ab initio* Mulliken analysis (in italic) are shown for each atom.